\documentclass[aps,prd,reprint,groupedaddress,amsmath,amssymb]{revtex4-2}
\usepackage{graphicx}
\usepackage{bm}
\usepackage{amssymb}
\usepackage{amsmath}
\usepackage{epsfig}
\usepackage{float}
\usepackage{makecell}
\usepackage{xcolor}
\usepackage{mathtools}
\usepackage{cases}
\usepackage{booktabs}

\usepackage{dcolumn}
\usepackage{bm}
\usepackage{braket}

\usepackage[
colorlinks=true,
filecolor=black,
anchorcolor=blue,
linkcolor=blue,
citecolor=magenta, 
urlcolor=magenta,
linktocpage=true,
plainpages=false,
breaklinks=true,
pdfstartview=FitH
]{hyperref}

\begin{document}
\title{
Parameter extraction of the stochastic gravitational wave background\\ with peak-like templates in millihertz}
\author{Heng-Sen Jiao$^{1,2}$}
\email
{jiaohs@bao.ac.cn}
\author{Hong-Bo Jin$^{1,2,4}$}
\email
{hbjin@bao.ac.cn}
\author{Yun-Long Zhang$^{1,3}$}
\email
{zhangyunlong@nao.ac.cn}

\affiliation{$^{1}$National Astronomical Observatories, Chinese Academy of Sciences, Beijing, 100101, China}

\affiliation{$^{2}$School of Astronomy and Space Science, University of Chinese Academy of Sciences, Beijing 100049, China.}
\affiliation{$^{3}$School of Fundamental Physics and Mathematical Sciences, Hangzhou Institute for Advanced Study, University of Chinese Academy of Sciences, Hangzhou 310024, China}
\affiliation{$^{4}$The International Center for Theoretical Physics Asia-Pacific (ICTP-AP), University of Chinese Academy of Sciences, Beijing 100190, China}

\begin{abstract}
We investigate a framework for extracting parameters of stochastic gravitational wave background (SGWB) with peak-like templates in the millihertz frequency band, and analyzing transient contamination effects on parameter reconstruction. We present the spectrum and spectrogram under different conditions and provide the results of parameter reconstruction. Using templates from the early universe, we demonstrate that the peak-like templates outperform the broken power law (BPL) templates in power-law exponents recovery and peak frequency localization. The reconstruction results obtained using data from Fast Fourier Transform (FFT) are better than those obtained using data from Short-Time Fourier Transform (STFT) which is based on the spectrogram. For the single-peak template, the estimation accuracy of the exponent and peak frequency surpasses that of the BPL template by an order of magnitude, but demonstrates less precision in amplitude estimation compared to BPL. Regarding the double-peak template, parameter estimation results derived from the FFT methodology consistently outperform those obtained using STFT. Nevertheless, transient signals exhibit a detrimental impact on parameter estimation precision, causing errors to increase by an order of magnitude, particularly in multi-peak scenarios. This framework provides an example for using templates to analyze data from space-based gravitational wave detectors.
\end{abstract}
\maketitle

\tableofcontents


\section{Introduction}
\label{sec1}

The Laser Interferometer Gravitational Wave Observatory (LIGO) is designed to detect gravitational waves (GW) from distant astrophysical sources in the 10 Hz to 10 kHz frequency range. Since the first detection of GW from merging solar-mass black holes in 2015~\cite{LIGOScientific:2016aoc}, a new observational window for the universe has been opened. In the nanohertz frequency range, Pulsar Timing Arrays (PTAs) serve as the primary detection method. PTA collaborations have reported evidence for a stochastic gravitational wave background (SGWB) in the nanohertz range~\cite{Agazie_2023}. In the 2030s, the Laser Interferometer Space Antenna (LISA)~\cite{Karnesis:2022vdp}, Taiji~\cite{Hu:2017mde}, and TianQin~\cite{Li:2024rnk} will open new windows into the GW spectrum in the millihertz range. Numerous signals are expected in this frequency range, including binary merger signals and various theoretically predicted GW signals, particularly gravitational wave backgrounds.

The SGWB has both astrophysical and cosmological origins, arising from numerous weak, independent, and unresolved sources. These astrophysical GW sources originate from various astronomical events, including binary mergers~\cite{GW170104,GW170608,GW170814,gw170817}, supernova explosions~\cite{Buonanno_2005,Yakunin_2010}, Extreme Mass Ratio Inspirals (EMRIs)~\cite{Gair_2017}, and other phenomena. As a key scientific target for space-based GW detectors, EMRIs have prompted studies on distinguishing their signals from others~\cite{Tanaka:2008zze} and identifying their event properties~\cite{Gair_2005}. In contrast to astrophysical GW sources, cosmological GW sources may reveal insights into the early universe and particle physics. Furthermore, GW observations provide information extending beyond these two domains~\cite{LISACosmologyWorkingGroup:2022jok}. For instance, supernova explosions are termed ``standard candles"~\cite{Hillebrandt_2000} due to their absolute magnitudes, while compact binary coalescences (CBCs) can be regarded as "standard sirens". The low-redshift measurement of the Hubble constant exemplifies application of this principle~\cite{Hubb2017}.

Although the SGWB exhibits significantly lower amplitudes than common astronomical events, it encodes crucial cosmic evolution information as a distinct noise component in GW signals. Distinguishing SGWB signal mixtures is vital, as physical origin insights depend on identifying their characteristic features. To effectively separate these signals, systematic methodologies are required. When disregarding specialized sources, a unified framework has been proposed in~\cite{Romano:2016dpx}. SGWB holds invaluable insights into the early universe, with numerous studies attempting its detection through LIGO and PTAs. Various works have imposed constraints on models including First-Order Phase Transitions (FOPTs)~\cite{Yu:2022xdw,Jiang:2022mzt}, cosmic strings in different frameworks~\cite{nielsen1973vortex,Dvali_2004,Copeland_2004,Chen:2022azo}, pre-Big Bang scenarios~\cite{Gasperini:1992em,Conzinu:2024cwl}, inflationary models~\cite{Starobinsky:1979ty,Grishchuk:1974ny,Easther:2006gt}, and reheating mechanisms~\cite{Turner:1993vb,Nakayama:2008wy}, etc.

In this paper, we focus on extracting SGWB signals with double-peaked frequency spectra, analyzing their spectrogram characteristics, and evaluating parameter estimation methods under varying noise conditions.
Spectrograms represent a widely adopted technique for recording gravitational wave events in detectors like LIGO. This valuable tool enables clear visualization of GW data, facilitates real-time analysis, and preserves the capability to reconstruct original signal features. In particular, binary merger signatures are prominently identifiable in spectrogram representations. The \texttt{SGWBinner} toolkit has been enhanced with a validated module to detect GW emissions from various cosmological sources \cite{Caprini:2024hue,LISACosmologyWorkingGroup:2024hsc,Blanco-Pillado:2024aca}.
We systematically categorize spectrogram features originating from GW sources with distinct formation backgrounds and transient signals. Central to our investigation is assessing the extract ability of these spectrogram features and the accuracy of original data reconstruction. In addition, we incorporate amplitude modulation effects with random jitter into our analysis framework. 

Transient GW signals, such as binary merger signals, are among the major sources detected by space-based gravitational wave interferometers. It is crucial to assess whether these transient signals affect the detection of the SGWB. Instrumental glitches are also considered a type of transient signal. Ref.~\cite{Iwanaga:2025muq} have quantified the impact of transient glitches on parameter estimation using machine learning. We implement parameter estimation protocols to characterize original spectral parameters and quantitatively evaluate the interference effects of transient signals on estimation accuracy.

This paper is organized as follows. In Sec.~\ref{sec2}, we introduce the approximate templates of GWs from the early universe.
In Sec.~\ref{sec3}, we demonstrate the characteristics of different templates and reconstruct the original spectrum from simulated data.
In Sec.~\ref{sec4}, we present the results of parameter reconstruction from simulated data and include supplementary corner plots in the Appendix \ref{app2}.

\section{GW Templates from the Early Universe}
\label{sec2}

The spectral shape provides critical information to distinguish the origin of the SGWB, and obtain further information about its underlying mechanism. Different astrophysical and cosmological production mechanisms produce characteristic spectral signatures determined by their respective physical processes. Future gravitational wave detectors can reconstruct the spectral shape by employing peak-like and broken power-law templates. We focus on the frequency range relevant to space-based gravitational wave detectors and evaluate the parameter space that can be probed by these instruments.

The energy density of GW is expressed as:
\begin{align}
\rho_{\rm GW}&= \frac{1}{64\pi G} \bigg\langle(\partial_t h_{ij})^2 +(\frac{\nabla}{a}h_{ij})^2 \bigg\rangle,\label{rho}
\end{align}
where $G$ is the gravitational constant, $a$ is the cosmological scale factor, $\nabla$ denotes the spatial derivative operator. The angular brackets indicate spatial averaging, assuming a flat Friedmann-Lemaître-Robertson-Walker (FLRW) universe. The metric perturbation $h_{ij}$ satisfies the transverse-traceless gauge conditions $\partial^i h_{ij}=h^i_{~i}=0$.
To characterize the spectral amplitude of GW, a dimensionless quantity $\Omega_{\rm GW}$ is defined to represent the energy density per logarithmic frequency interval:
\begin{align}
\Omega_{\rm GW}(f)  &=  \frac{1}{\rho_{crit}} \frac{d\rho_{GW}}{d\ln{f}},\label{Omegagw}
\end{align}
where $\rho_{crit} = 3H_*^2/8\pi G$, $H_*$ is the Hubble constant at the time when GW generated.

In the following, we will introduce peak-like templates and broken power-law templates, both of which are employed to model the GW spectrum of the SGWB.

\subsection{Peak-like templates}
The single-peak (SP) template is employed to characterize the spectral energy distribution of a singular SGWB source from the early universe. This template takes the  form:
\begin{equation}
\Omega_{\rm SP}(f) =
\begin{aligned}
\begin{cases} 
\Omega_{*} \left( \frac{f}{f_{*}} \right)^{n_{1}}, & \text{} f < f_{*}, \\
\Omega_{*} \left( \frac{f}{f_{*}} \right)^{n_{2}}, & \text{} f > f_{*},
\end{cases}
\end{aligned}
\label{eq1}
\end{equation}
Here, $n_{1}$ and $n_{2}$ are power-law exponents, typically satisfying $n_1>0$ and $n_2<0$.  The quantity $\Omega_{*}$ denotes the amplitude at the peak frequency $f = f_{*}$ (or equivalently,  characteristic frequency).
This approximate template applies to most SGWB scenarios and facilitates a fundamental understanding of their key spectral features.
In Eq.~(\ref{eq1}), we specify the parameters for the SP template as follows:
\begin{equation}
\begin{aligned}
n_1 = 3,~ n_2=-4, ~f_{*}=1\text{mHz},~\Omega_*=7.62 \times 10^{-10}. \label{parametersp}
\end{aligned}
\end{equation}
 
The double-peak (DP) template is used to describe scenarios in which the spectra of multiple sources are superimposed. We formalize the DP case as follows:
\begin{equation}
\Omega_{\rm DP}(f) =
\begin{aligned}
\begin{cases} 
\Omega_{1} \left( \frac{f}{f_{1}} \right)^{n_{1}} + \Omega_{2} \left( \frac{f}{f_{2}} \right)^{n_{3}}, & \text{} f < f_{1}, \\
\Omega_{1} \left( \frac{f}{f_{1}} \right)^{n_{2}} + \Omega_{2} \left( \frac{f}{f_{2}} \right)^{n_{3}}, & \text{} f_{1}< f < f_{2},\\
\Omega_{1} \left( \frac{f}{f_{1}} \right)^{n_{2}} + \Omega_{2} \left( \frac{f}{f_{2}} \right)^{n_{4}}, & \text{} f > f_{2}.
\end{cases}\label{2p}
\end{aligned}
\end{equation}
Here, $\Omega_1$ and $\Omega_2$ represent the amplitudes of the first and second peaks, respectively. The quantities
$n_i$ denote the power-law exponents, with typical constraints $n_1>0,n_3>0$ and $n_2<0,n_4<0$. 
We collectively refer to SP and DP templates as peak-like templates.
According to \cite{2018Probing}, we select the following parameter combination for the double-peak template in Eq.~\eqref{2p}:
\begin{equation}
\begin{aligned}
&n_1 = 2.85,~n_2=-2.6, ~f_1=1 \text{mHz},~\Omega_1= 1.02\times 10^{-9},  \\
&n_3 = 3,\quad~~ n_4=-4,\quad f_2=2 \text{mHz},~ \Omega_2= 7.62 \times 10^{-10}. \label{parameter}
\end{aligned}
\end{equation}
 
For the first peak, one typical GW source can originate from inflationary processes, such as the second slow-roll phase and hybrid inflation with a mild waterfall stage. Based on comprehensive analysis, we determine the allowable parameter ranges for both models. The resulting spectrum from this parameter combination aligns with the parameter constraints outlined in~\cite{LISACosmologyWorkingGroup:2024hsc}.
For the second peak, we adopt the sound wave mechanism from FOPT, as elaborated in the Appendix \ref{app1}. Both the Electro-Weak Phase Transition and the Quantum Chromodynamic Phase Transition may produce GWs detectable by space-based gravitational wave detectors \cite{Apreda:2001tj,Apreda:2001us,Grojean:2006bp}. FOPT releases energy with such violence that it generates GWs observable today.

The peak-like templates can effectively characterize GW spectra originating from different physical mechanisms, as they require only adjustments to power-law exponents and peak frequencies. Various physical processes that occurred in the early universe produce correspondingly different spectral signatures. A comprehensive overview of the template parameters for various SGWB sources can be found in~\cite{2018Probing}.

\subsection{Broken power-law templates}

In contrast to the peak-like templates discussed above, the broken power-law spectrum offers greater flexibility in controlling the rate of the transition from low to high frequency. It can be modeled as:
\begin{equation}
\Omega_{\rm BPL}(f) = \Omega_{*} \left(\frac{f}{f_*}\right)^{n_1}\left[\frac{1+(f/f_*)^{\Delta}}{2}\right]^{(n_2-n_1)/\Delta} \label{bpl},
\end{equation}
where $\Omega_*$ represents the peak amplitude at the characteristic frequency $f_*$. The parameters $n_1$ and $n_2$ denote the asymptotic spectral indices for the low-frequency and high-frequency tails, respectively. 
$\Delta^{-1}$ characterizes the transition width between these asymptotic regimes. 
For the BPL template, we adopt parameter values within the permissible range established in the relevant literature~\cite{LISACosmologyWorkingGroup:2024hsc}:
\begin{equation}
\begin{aligned}
&n_1=2.85,\quad n_2=-2.6,\quad \Delta=1,\\
&f_{*}=2\text{mHz},\quad \Omega_{*}=6.97\times10^{-10}.\label{parameterbpl}
\end{aligned}
\end{equation}

The BPL template enables us to model physical phenomena with enhanced accuracy. By adjusting the parameter $\Delta$ in Eq.~\eqref{bpl}, the spectral descent can be regulated more smoothly, thereby avoiding abrupt reductions to zero. This parameter-controlled modulation of descent rate facilitates more realistic modeling of diverse physical phenomena, particularly those where high-frequency components exhibit gradual decay governed by underlying physical mechanisms, rather than demonstrating sharp spectral cut-offs.

\section{Signal Characteristic Extraction}
\label{sec3}

In this section, we investigate the spectral characteristics of distinct signals within the spectrogram framework and reconstruct their frequency-domain representations from synthetic datasets. Our analysis of the GW spectrum employs both the DP template defined in Eq.~\eqref{2p} and the BPL template from Eq.~\eqref{bpl}. For transient signal morphologies, the two templates exhibit negligible differences in their injection characteristics; consequently, the DP template in Eq.~\eqref{2p} is adopted for subsequent analytical procedures. The systematic effects of synthetic data on parameter estimation will be discussed in Section~\ref{sec4}.

\subsection{Methodology for signal analysis}

Given that the background signal exhibits time independence, whereas transient signals manifest over limited temporal intervals, a standard spectrogram suffices to visualize their spectral characteristics. For signal analysis, we employ both the FFT and Short-Time Fourier Transform STFT methodologies.
FFT represents an efficient algorithmic implementation of the Discrete Fourier Transform (DFT), fundamentally designed for computational efficiency. The STFT methodology, in contrast, operates by systematically applying DFT to segmented signal portions within sliding time windows, thereby capturing time-varying spectral features.

For DFT, the mathematical expression is as follows:
\begin{align}
    H[k]&= \sum^{N-1}_{n=0}h[n] \cdot e^{-i \frac{2\pi k n}{N}},\\
    h[n]&= \frac{1}{N} \sum^{N-1}_{k=0} H[k]\cdot e^{i \frac{2\pi k n}{N}},
\end{align}
where $N$ is the length of the signal, $n$ denote the time points. $k$ denote the frequency values. $h[n]$ is the time-domain data and $H[k]$ contains the frequency domain information of amplitude. While DFT directly computes frequency components with $O(N^2)$ complexity, FFT exploits symmetry and periodicity in the DFT matrix to reduce the number of operations to $O(N \text{log}N)$.

For STFT, we use the discrete form to describe the mathematical expression:
\begin{align}
    H[m,k]=\sum^{N-1}_{n=0} h[n] \cdot \mathcal{W}[n-mL] \cdot e^{-i \frac{2\pi k n}{N}}, \label{stft}
\end{align}
where $m$ denotes the time point, $L$ is the hop size of the window function. We can analyze signals in different aspects by adjusting the length and form of the window function. To effectively reduce spectral leakage, Hamming window function is selected for our study:
\begin{equation}
    \mathcal{W}(n) = \frac{1}{2}\left(1- \cos{\frac{2\pi n }{N}}\right).
\end{equation}
During the process of reconstructing the origin spectrum, slight changes in amplitude and phase may occur during processing, but these changes have limited influence on the identification and reconstruction of key features in the spectrogram.

\begin{figure}[h]
    \centering
    \includegraphics[width=0.75\linewidth]{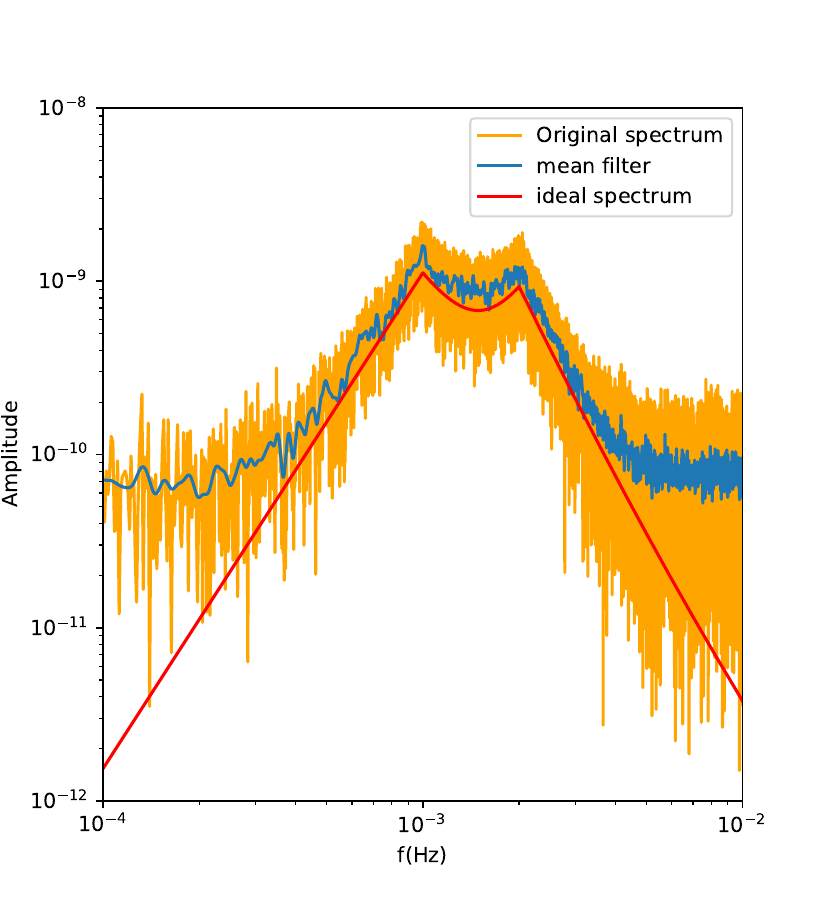}
    \caption{The spectrum is obtained by processing time-series data with random phase, noise, and random jitter. The orange line is obtained directly by applying FFT to the time-series data, and the blue line is generated by applying mean filtering to the orange line. The red line is derived based on the DP template in Eq.~(\ref{2p}) and parameters in Eq.~(\ref{parameter}).}
    \label{fig1}
\end{figure}

Extracting the true signal requires characterizing the noise properties (e.g., Gaussian, colored, or non-stationary) to choose an optimal filter. The matched filtering is designed to maximize the signal-to-noise ratio by using an expected template to filter out noise that does not match this template \cite{Renzini:2023qtj}. This approach is particularly effective when the noise is distributed over a broad frequency range and exhibits significant amplitude variations. In the case of highly complex noise models, which are common in space-based gravitational wave interferometers, advanced techniques like matched filtering are required. The ideal waveform in the original spectrum from Fig.~\ref{fig1} is obscured by noise. 

Besides, the constant Q spectrogram is widely used in LIGO \cite{LIGOScientific:2016aoc} and future GW detection \cite{Abac:2025saz}, because a wide window function is required for low-frequency time-varying signals, while a narrow window is required for high-frequency time-varying signals due to their rapid frequency conversion.
Such an approach is particularly crucial for studying transient signals, such as those produced by binary black hole mergers, where the frequency content changes dynamically as the system progresses. Examining the time-frequency evolution enables the identification of key signal features, including frequency shifts, amplitude variations, and transient behaviors, all of which are critical for precise detection and characterization of astrophysical events.
To improve the parameter estimation of the mock data, we will set the width of the window function to a fixed value in next subsection.

\begin{figure}[ht]
    \centering
\includegraphics[width=1.05\linewidth]{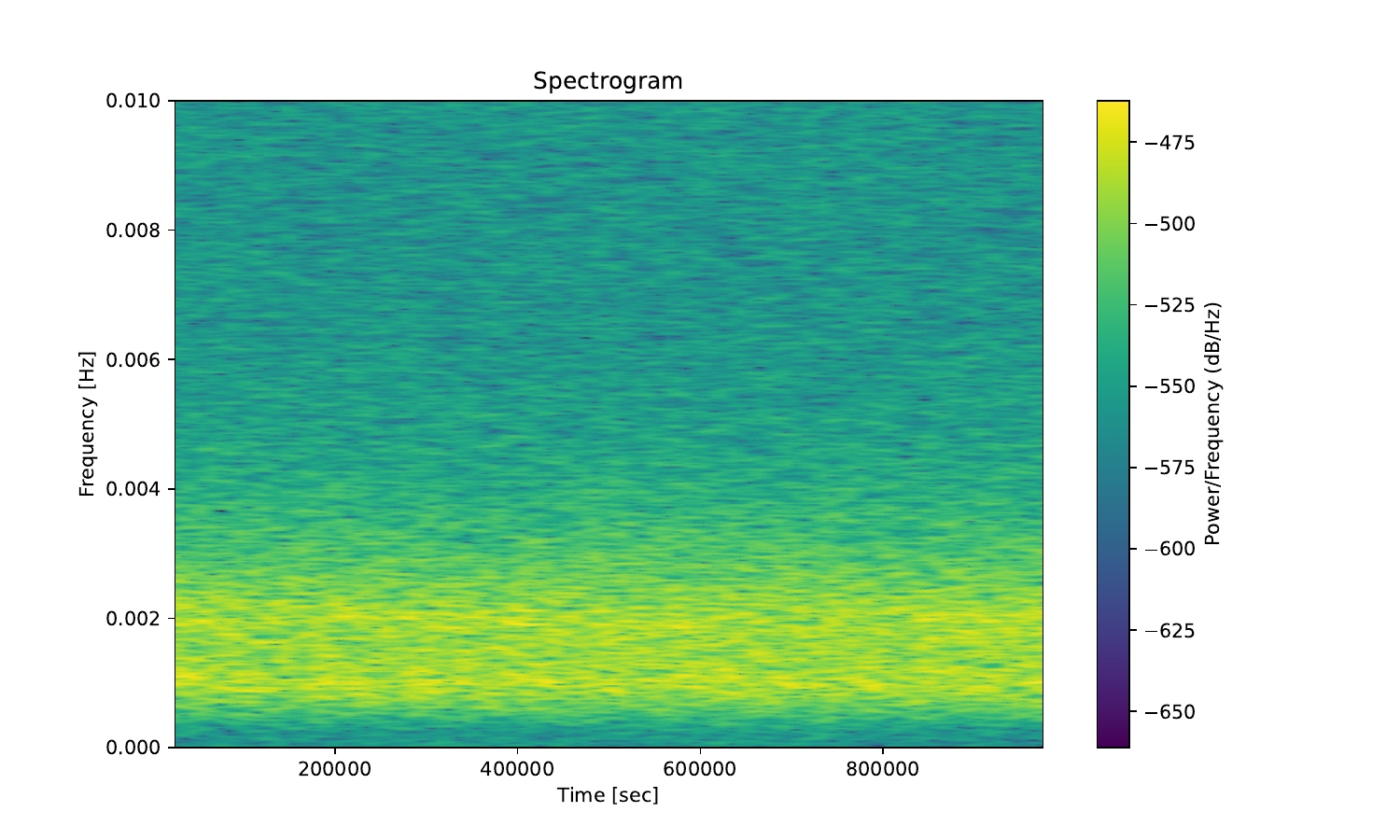}
    \caption{The spectrogram is obtained from the time series data generated by the original spectrum from Fig.~\ref{fig1}. The two yellow region denote two peak in 1mHz and 2mHz respectively, with the DP template in Eq.~(\ref{2p}) and parameters in Eq.~(\ref{parameter}).}
    \label{fig4}
\end{figure}

\subsection{Spectral reconstruction}
\label{sec3b}

In this subsection, we begin with spectral reconstruction based on the time-series mock data. After constructing mock data, we show their characteristics in spectrogram. 
Theoretical analyses typically assume consistent phase or predetermined phase evolution in ideal signals, thus clearly revealing characteristics. However, in practical scenarios, this assumption is unrealistic. The phase is typically random due to various physical and environmental factors \cite{LIGOScientific:2019vic}, necessitating consideration of this randomness in mock data. 

We consider the signal with both random phase and noise added simultaneously and show its spectrogram in Fig.~\ref{fig4}. From the spectrograms we can estimate the approximate energy distribution. The frequency range containing the peak can be identified. To determine the specific source, filtering and parameter estimation must be applied to these data for further reconstruction.

\begin{figure}[ht]
    \centering
    \includegraphics[width=0.85\linewidth]{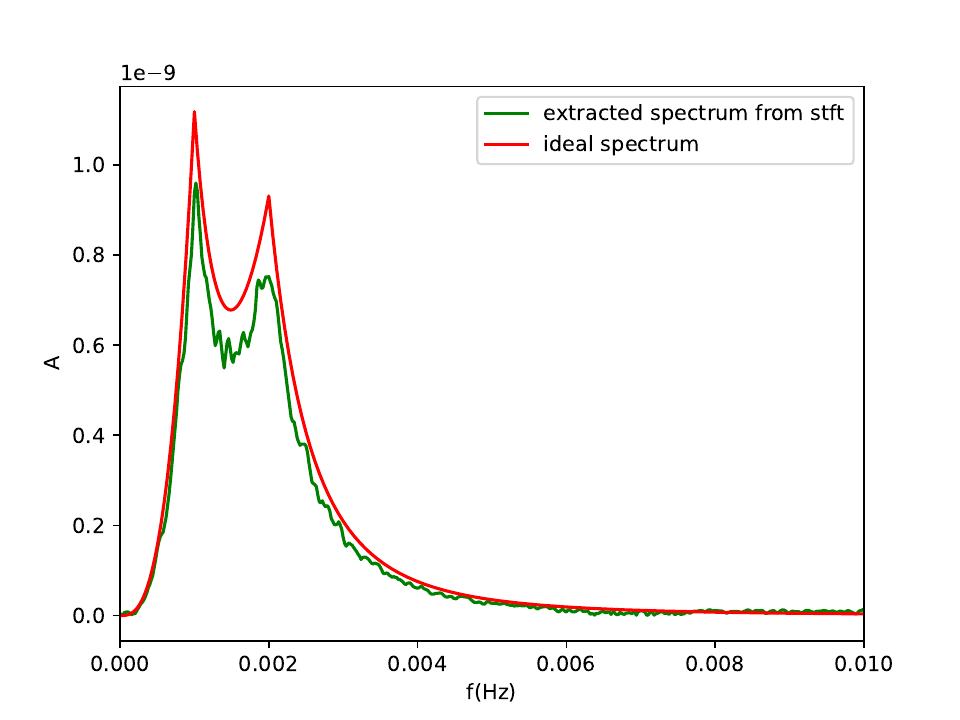}
    \caption{The red line is the ideal spectrum from the DP template in Eq.~(\ref{2p}) and parameters in Eq.~\eqref{parameter}, and the green line is the spectrum reconstructed from Fig.~\ref{fig4} by using STFT and Eq.~(\ref{aver}). }
    \label{fig5}
\end{figure}

For the traditional inverse STFT, one considers not only the overlap points, but also the form of the window function. This complexity arises because the reconstruction process is highly sensitive to both the shape of window function and the overlap parameters. The original signal can only be reconstructed under specific window functions and conditions, and signal filtering remains a challenge. Here, we sum the spectra of each window function and then take the average to obtain the original spectrum:
\begin{equation}
    \Omega_{*}(f)=\frac{1}{N}{\sum_{m=1}^{N}\Omega_{}(m,f)}.\label{aver}
\end{equation}
This operation is physically reasonable, as it simplifies the process by discarding temporal dependencies and effectively capturing the global spectral characteristics.
After averaging the spectrum over different time windows, the characteristics become more apparent. At the very least, the peak frequency range can be obtained from the spectrum. It can be observed that the amplitude at the peak frequency differs, which is due to the inherent spectrum leakage of STFT. The Hamming window can make the spectrogram smoother and make it easier to visualize how the frequency evolves over time.

\begin{figure}[h]
\centering
\includegraphics[width=0.75\linewidth]{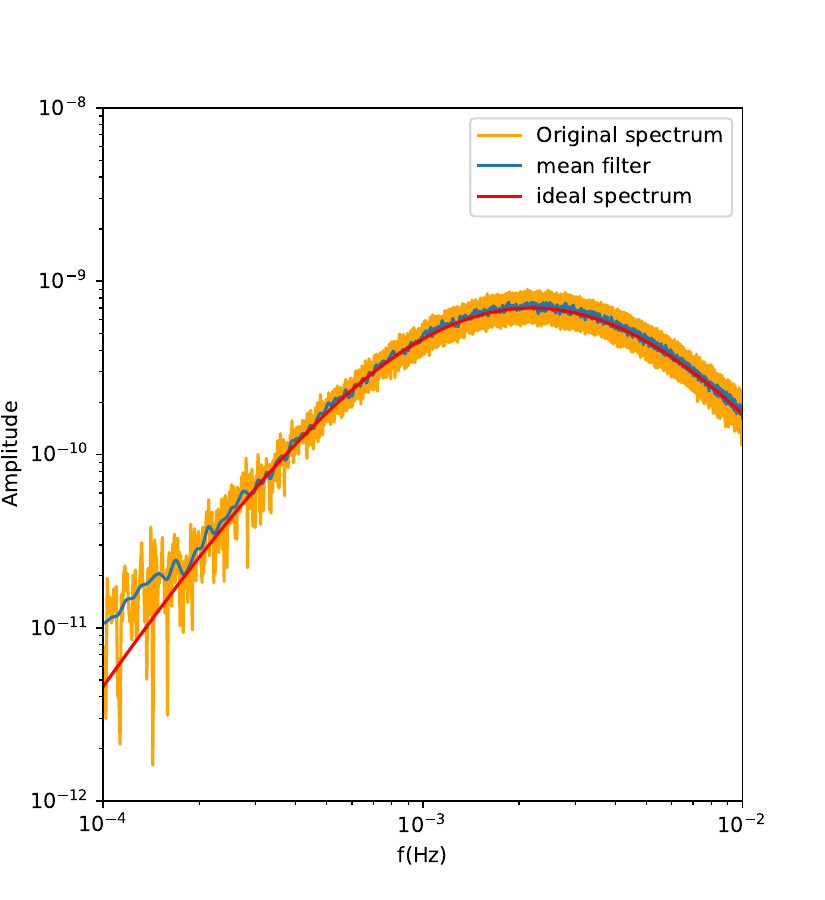}
\caption{The spectrum is obtained by processing time-series data with added random phase, noise and random jitter. The orange line is obtained directly by applying FFT to the time-series data. The blue line is generated by applying mean filtering to the orange line. The red line is derived based on the BPL template in Eq.~(\ref{bpl}) and parameters in Eq.~\eqref{parameterbpl}.}
    \label{fig8}
\end{figure}

\begin{figure}[h]
\centering
\includegraphics[width=1.05\linewidth]{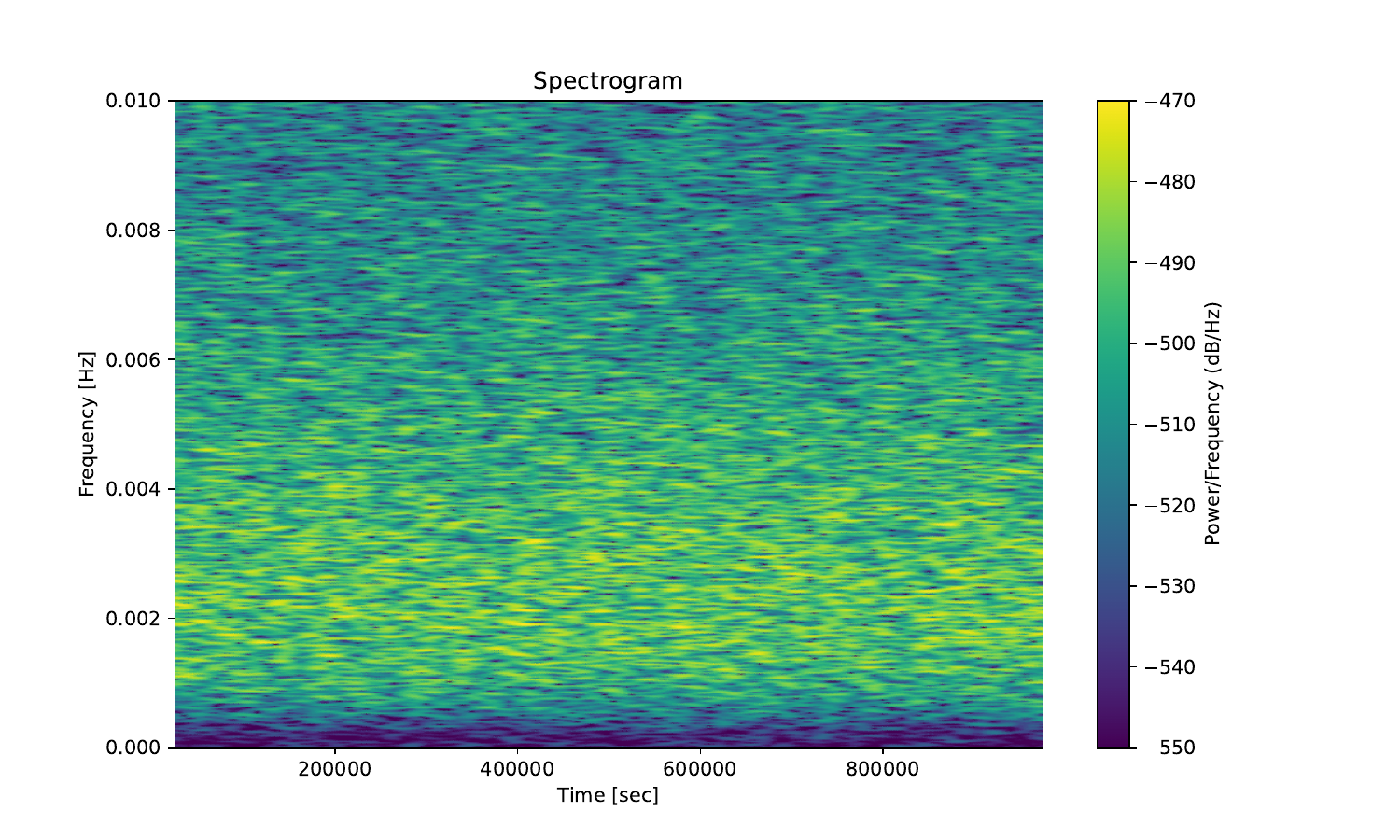}
\caption{The spectrogram is obtained from the time-series data generated by the original spectrum from Fig.~\ref{fig8}, which shows the concrete distribution of energy from BPL, with the BPL template in Eq.~(\ref{bpl}), and parameters in Eq.~\eqref{parameterbpl}.}
\label{fig9}
\end{figure}

From the results of STFT, it can be concluded that restoring the original spectral characteristics only through FFT and STFT is challenging without applying the filter method and noise reduction. For Fig.~\ref{fig5}, we have reduced the noise, applied a filter method, and reconstructed the spectrum affected by the window function. 
In this case, finding an effective method for spectrum reconstruction is crucial. Otherwise, it is difficult to identify the background signal. In fact, even white noise significantly lower than the peak amplitude can severely interfere with spectrum reconstruction.

\begin{figure}[h]
    \centering
\includegraphics[width=0.85\linewidth]{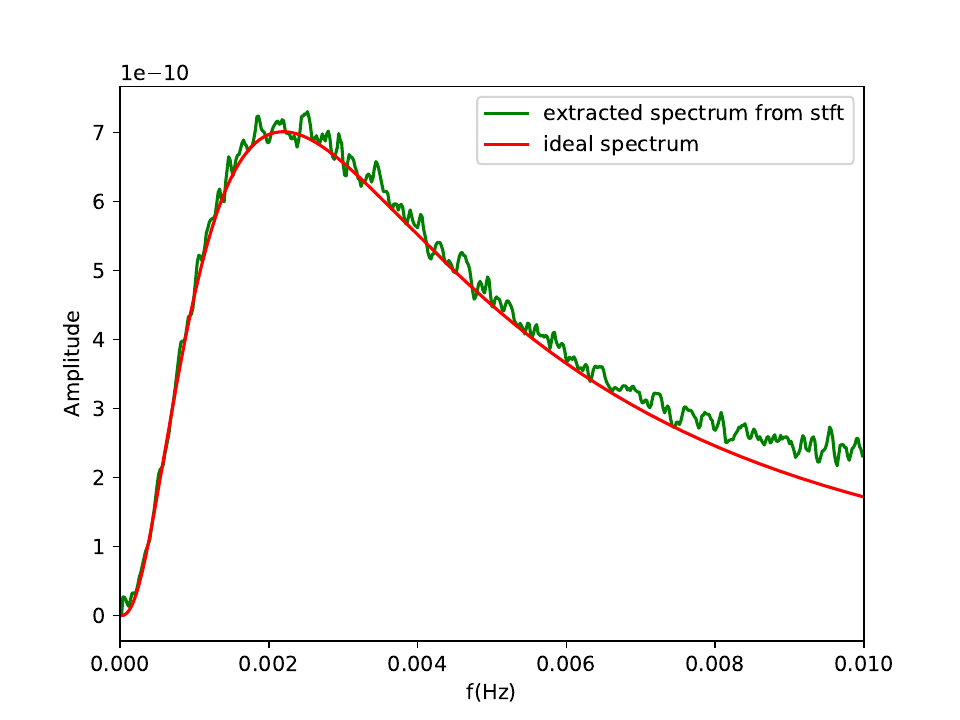}
    \caption{The red line is the ideal spectrum from the BPL template in Eq.~(\ref{bpl}) and parameters in Eq.~\eqref{parameterbpl}. The green line is the spectrum reconstructed from the spectrogram in Fig.~\ref{fig9} by using Eq.~(\ref{aver}).}
    \label{fig9+1}
\end{figure}

As a comparison, we use the same methods to generate the mock data and reconstruct the spectrum based on the BPL template in Eq.~(\ref{bpl}). The results are shown in Fig.~\ref{fig8}, Fig.~\ref{fig9}, and Fig.~\ref{fig9+1}.

\subsection{Effects of transient signal} \label{Sec3C}

To investigate methods for distinguishing transient signals from the SGWB, we inject a segment of binary merger signals into noise-contaminated signals. The merger signal can be approximately expressed as~\cite{Blanchet:2013haa,Grishchuk:1988kq}:
\begin{equation}
h(t)=A(t) \sin {\phi(t)}, \label{merger signal}
\end{equation}
where
\begin{align}
A(t) &= A_0 [f(t)]^{\frac{2}{3}},~~
 f(t)=f_0 (t_{0}-t)^{-\frac{3}{8}},\\
\phi(t) &= \int{2\pi f(t)}dt=\frac{16\pi}{5}  f_0  (t_{0}-t)^\frac{5}{8}.
\end{align}
We will choose the parameters as follows:
$f_0 = 0.6\text{mHz}$,
$A_0 =5\times 10^{-12}$,
and $t_{0}$ is the time when the merger occurred.

\begin{figure}[h]
\centering
\includegraphics[width=1.05\linewidth]{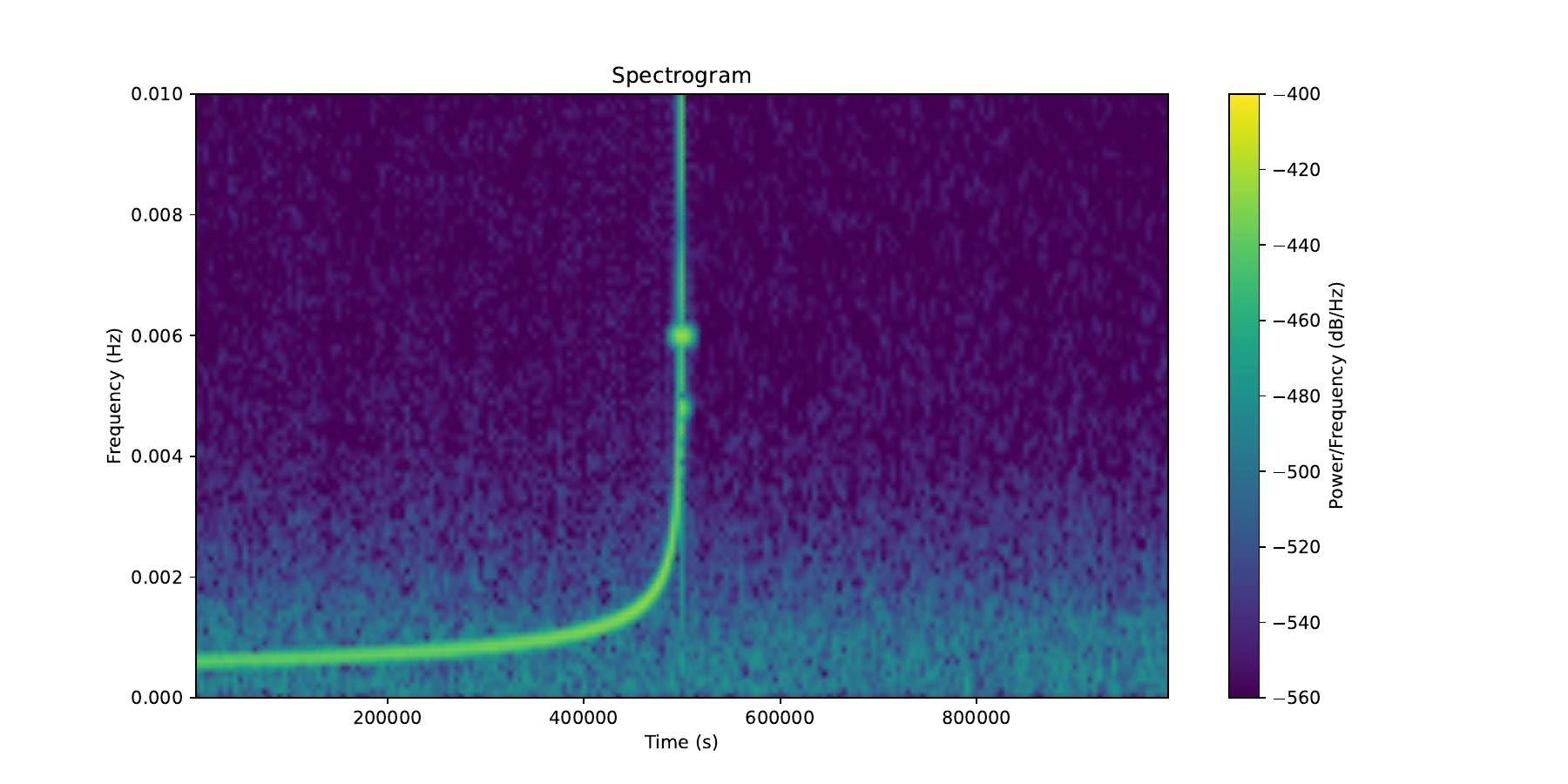}
    \caption{SGWB spectrogram with the DP template in Eq.~(\ref{2p}), parameters in Eq.~(\ref{parameter}) and
    a segment of binary merger signal in Eq.~\eqref{merger signal}.}
    \label{fig6}
\end{figure}

Fig.~\ref{fig6} clearly shows a rapid increase in signal intensity from the low-frequency region to the relatively high-frequency region. To investigate the impact of transient signals on both the spectrum and parameter estimation, we average the data over a given period to restore the spectrum. By applying the STFT results, we reconstruct the signal using Eq.~(\ref{aver}). The transient signal appears at 6 mHz. Our aim is to explore how to detect this signal when applying STFT. To optimize the visibility of the signal on the spectrogram, we adjust the STFT window function and resolution parameters. This adjustment allows us to examine how these parameters influence the resolution of transient signals.

\begin{figure}[h]
\centering
\includegraphics[width=0.85\linewidth]{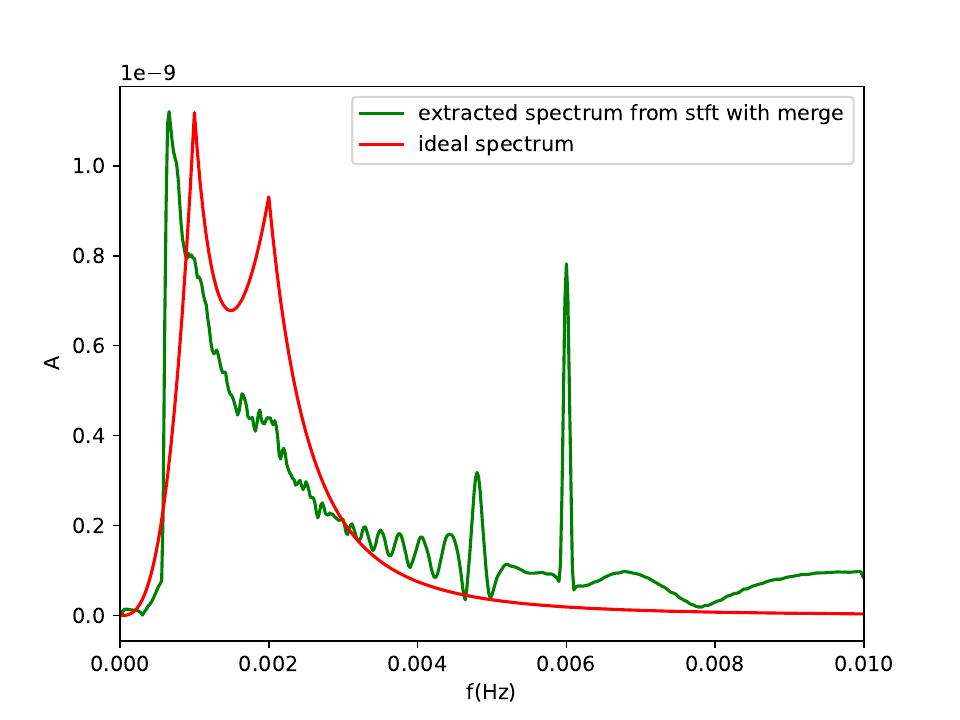}
\caption{Comparison between ideal spectrum and extracted spectrum from STFT data with transient signal, based on the DP template in Eq.~(\ref{2p}) and parameters in Eq.~\eqref{parameter}.}
    \label{fig7}
\end{figure}

Since the amplitude of a binary merger signal varies with frequency, combining the spectrogram in Fig.~\ref{fig6}, which provides time-dependent frequency content, and the spectrum in Fig.~\ref{fig7}, which captures overall frequency distribution, provides a more complete understanding of the signal.
By integrating these two representations, we can track the evolution of signal over both time and frequency domains, yielding a more comprehensive understanding of its behavior. This dual-domain analysis is essential for accurately analyzing and detecting GWs, as it captures time-dependent variations in frequency and amplitude.

From the extracted spectrum in Fig.~\ref{fig7}, it is observable that the two peaks at 1mHz and 2mHz become indistinguishable due to the presence of transient signals. This phenomenon arises because the energy of the transient signals is concentrated around 6mHz, generating a new peak at 6mHz. Simultaneously, the inspiral stage causes the original two peaks to merge.
Another issue with the peak-like template is its dense structure, which makes recognizing and distinguishing the background signal extremely challenging. This underscores the need for caution when removing transient signals, as they can significantly impact the accuracy of the analysis. Accurately reconstructing the original peak spectrum is key to both feature identification and parameter reconstruction.

In addition, after injecting the same signal into both templates in Eq.~\eqref{2p} and Eq.~\eqref{bpl}, the extracted results reveal additional issues. For the peak-like template in Fig.~\ref{fig7}, traces of the original background signal are scarcely detectable in the spectrum. We might even infer that the second peak is at 6mHz, given its amplitude of approximately $8\times10^{-10}$. Thus, using an incorrect spectrum for parameter estimation leads to a very large error in the parameters. This indicates that the peak-like template may be significantly influenced by noise and transient signals. Consequently, the background signal of the peak virtually disappears, posing significant challenges for subsequent analysis and signal detection, particularly when addressing future binary merger signals.

\section{Parameter Reconstruction} \label{sec4}

In this section, we employ multi-parameter Bayesian estimation for parameter reconstruction. In a simplified model-independent scenario, random phase and white noise are added to the signal to represent measurement errors. To prevent errors arising from floating-point precision during sampling, we utilize amplified values for sampling, which theoretically do not affect the correctness of the parameter estimation.

\subsection{Parameter estimation}

Parameter estimation serves as a powerful tool for reconstructing parameters related to GW events~\cite{Lyu:2023ctt,Gao:2024uqc}. In Appendix \ref{app2}, we present the parameter estimation results in figures, with likelihood functions derived from FFT and STFT, respectively. Specifically, Fig.~\ref{fig:single} presents the corner plot for the SP template defined in Eq.~(\ref{eq1}), which includes parameters initialized according to Eq.~\eqref{parametersp}. Fig.~\ref{fig:11} and Fig.~\ref{fig:12} present corner plots for the BPL template in Eq.~(\ref{bpl}), demonstrating the comparison of parameter reconstruction accuracy between FFT-derived and STFT-derived data, respectively.

A comparison of the SP template in Fig.~\ref{fig:single} and the BPL template in Fig.~\ref{fig:11} reveals that the amplitude of SP template exhibits lower accuracy. This discrepancy arises because the peak in SP template is highly sensitive to random jitter, defined as a percentage, which induces a larger absolute error in the peak when compared to the BPL template. Consequently, parameter reconstruction for these two templates demonstrates distinct applicability depending on specific requirements. For the SP template, we obtain a more accurate value of the peak frequency $f_{*}$, whereas the BPL template results in a higher absolute error. This is attributed to the fact that the descent of the spectrum can be modulated more smoothly by $\Delta$  in the BPL template. The latent physical mechanism underlying these templates leads to different parameter values, and selecting appropriate models allows us to extract relevant physical information.

For the DP template in Eq.~\eqref{2p}, we utilize parameter combinations from Eq.~\eqref{parameter} for sound waves in FOPT and inflationary to investigate noise influence on the model parameters~\cite{Caprini:2024hue,LISACosmologyWorkingGroup:2024hsc}. As previously discussed, for noisy signal segments, applying a FFT to the entire signal followed by filtering generally yields better performance than averaging via the STFT method. This advantage arises because the Fourier transform preserves the global frequency content, avoiding the information loss caused by STFT windowing. Moreover, the filtering process is more effective at isolating relevant features when the entire frequency spectrum is considered as a whole. To validate this observation, we conduct parameter estimation using both methods independently, to enable a systematic comparison of their effectiveness under various scenarios. This comparison also provides insights into the trade-offs between time resolution and frequency resolution inherent in these two approaches.

In Appendix \ref{app2}, Fig.~\ref{fig10}, Fig.~\ref{fig11}, and Fig.~\ref{withmerge} show the corner plots for the DP template given by Eq.~(\ref{2p}). Specifically, the first two figures validate parameter estimation accuracy using data from FFT and STFT respectively, while the third explicitly quantifies the errors introduced by merger-related signals in the parameter reconstruction process. These corner plots demonstrate the accuracy of parameters across different templates, with lower relative errors indicating higher result accuracy.
For the histograms in these plots, median values and the  $2\sigma$ confidence level (95\%) are marked by blue dashed lines, whereas the blue solid line represents the true input value. 
Unlike $\sigma$ in a one-dimensional normal distribution, the two-dimensional distribution contains sample proportions (11.8\%, 39.3\%, 67.5\%,  86.4\%) corresponding to (0.5$\sigma$, 1$\sigma$, 1.5$\sigma$, 2$\sigma$).These four contour lines in the contour plot correspond to the specified confidence levels.

\subsection{Parameter comparison}

From the parameter estimation, using only FFT results yields a narrower parameter range for determination in Fig.~\ref{fig10}, whereas STFT results produce a broader and less precise parameter range in Fig.~\ref{fig11}. The discrepancy arises primarily from two factors: First, under identical computational resources, FFT achieves higher likelihood accuracy through its superior global frequency resolution, whereas STFT offers comparatively fewer data points.Second, regarding window size optimization, excessively large windows fail to capture signal frequency variations, while overly small windows introduce deviations in frequency components. In the field of time-frequency analysis, this fundamental trade-off is known as the Gabor limit, representing the inherent uncertainty principle between temporal and frequency resolutions.

In addition, increasing the number of data points $N$ in STFT results in a computational cost scaling approximately as $N^2$, since each $m$ in Eq.~(\ref{stft}) requires FFT application. While effective for identifying transient sources, the substantial computational burden and compromised frequency resolution associated with STFT make it less suitable for high-precision parameter estimation.
To address this efficiency issue, we propose a simplified spectral reconstruction method. Specifically, we substitute the complex inverse STFT computation with Eq.~\eqref{aver}, achieving comparable accuracy while disregarding the specific window function formulation. This approach maintains minimal accuracy loss while significantly reducing computational complexity.

\begin{widetext}
\begin{center}

\begin{table}[ht]
\centering
\begin{tabular}{|c|c|c|c|c|c|} \hline 
Templates & $10^{10}\Omega_{*}$& $n_1$& $n_{2}$&$\Delta$ & $f_*/\text{mHz}$\\ \hline  
SP (FFT) & $9.76$\footnotesize\makecell{$+0.15$\\$-0.15$}& $3.02$\footnotesize\makecell{$+0.07$\\$-0.07$}& $-4.11$\footnotesize\makecell{$+0.12$\\$-0.13$} & $-$ & $2.01$\footnotesize\makecell{$+0.01$\\ $ -0.01 $}\\ \hline 
BPL (FFT) &$7.12$\footnotesize\makecell{$+0.19$\\$-0.42$}&$2.45$\footnotesize\makecell{$+1.10$\\$-0.70$}&$-2.39$\footnotesize\makecell{$+0.44$\\ $-0.73$} & $1.14$\footnotesize\makecell{$+0.43$\\ $ -0.33 $} & $2.19$\footnotesize\makecell{$+0.33$\\ $ -0.34 $}\\ \hline 
BPL (STFT) &$6.90$\footnotesize\makecell{$+0.30$\\$-0.44$}&$1.96$\footnotesize\makecell{$+0.72$\\$-0.55$}&$-1.57$\footnotesize\makecell{$+0.41$\\ $-0.70$}& $1.59$\footnotesize\makecell{$+0.85$ \\ $-0.52$} & $1.96$\footnotesize\makecell{$+0.52$\\  $-0.40$}\\ \hline 
\end{tabular}
\caption{Comparison of parameter estimation between SP template in Eq.~\eqref{eq1} and BPL template in Eq.~\eqref{bpl}. }
\label{table1}
\end{table}

\begin{table}[ht]
\centering
\begin{tabular}{|c|c|c|c|c|c|} \hline 
Templates & $\Omega_{*}$& $n_1$& $n_{2}$& $\Delta$&$f_*$\\ \hline  
SP (FFT) & $28.08\%$  & $0.67\%$ & $2.75\%$& $-$ & $  0.50\%     $ \\ \hline 
BPL (FFT) &$2.15\%$ & $4.01\%$  & $8.07\%$  & $14.00\%$ & $9.50\%$  \\ \hline 
BPL (STFT) &$1.00\%$ & $31.23\%$  & $39.62\%$  &  $59.00\%$ & $2.00\%$  \\ \hline 
\end{tabular}
\caption{Comparison of relative errors between the most probable values and initial values: for the SP template in Eq.~\eqref{eq1} using the initial parameters from Eq.~\eqref{parametersp}, and for the BPL template in Eq.~\eqref{bpl} using the initial parameters from Eq.~\eqref{parameterbpl}.}
\label{table3}
\end{table}

\end{center} \end{widetext}

Our comparative analysis of SP and BPL templates for parameter estimation, as presented in Tables~\ref{table1} and \ref{table2}, reveals distinct performance patterns across different parameters. For spectral index reconstruction, the SP template demonstrates superior precision in $n_1$ and $n_2$ estimates, respectively showing $0.67\%$ and $2.75\%$ relative deviations from injection values. This performance exceeds by an order of magnitude the $4.01\%$ and $8.07\%$ deviations observed in the BPL template. Conversely, the BPL template exhibits enhanced $\Omega_*$ estimation accuracy with a $2.15\%$ deviation, whereas the SP template shows a $28.08\%$ deviation. These findings underscore the parameter-specific efficacy of template selection in gravitational wave analyses, emphasizing the need for targeted template optimization based on individual parameters.

From Tables~\ref{table3} and \ref{table4}, the merger signal brings large error to the result of parameter estimation. Because this merger signal has a great impact on the spectrum. 
This indicates that the first step in analyzing SGWB and focusing on a specific signal segment, the first step is to determine whether transient signals are present. 
If there is a transient signal, the impact of the transient signal on the overall reconstruction needs to be considered. It can be divided into two situations.

Thus, transient signals introduce significant errors in parameter reconstruction. Their influence necessitates mitigation via time-frequency analysis prior to reconstruction.
If the transient signals do not significantly affect the research target, the signal segment can be extracted directly since the SGWB is generally stationary over time.
At a minimum, if the transient signals occupies only a small portion of the total time period under analysis, it can be reasonably ignored due to the time-averaging effect. This effect ensures that the contribution of transient signals becomes negligible when average over a sufficiently long time scale, minimizing their impact on the overall parameter estimation process.

\begin{widetext} \begin{center}

\begin{table}[H]
\centering
\begin{tabular}{|c|c|c|c|c|c|c|c|c|} \hline 
Templates & $10^{10}\Omega_{inf}$ & $10^{10}\Omega_{pt}$ & $n_{1}$ & $n_{2}$ & $n_{3}$ & $n_{4}$ &$f_1/{\rm mHz}$& $f_2/{\rm mHz}$    \\ \hline  
DP (FFT)& $11.48$\footnotesize\makecell{$+1.10$\\$-0.94$}& $9.67$\footnotesize\makecell{$+1.06$\\$-1.04$}& $3.08$\footnotesize\makecell{$+0.30$\\$-0.21$}&$-2.66$\footnotesize\makecell{$+0.57$\\$-0.98$}&$2.23$\footnotesize\makecell{$+0.94$\\$-0.43$}&$-4.17$\footnotesize\makecell{$+0.50$\\$-0.81$}
& $0.99$\footnotesize\makecell{$+0.01$\\ $-0.01$}& $2.03$\footnotesize\makecell{$+0.02$ \\$-0.02$}\\ \hline 
DP (STFT)&$7.14$\footnotesize\makecell{$+1.96$\\$- 1.79$}&$6.31$\footnotesize\makecell{$+1.35$\\$-1.64$}&$3.13$\footnotesize\makecell{$+2.52$\\ $-1.20$}&$-2.89$\footnotesize\makecell{$+1.33$ \\$-3.02$}&$1.73$\footnotesize\makecell{$+3.44$\\  $-0.65$}&$-4.20$\footnotesize\makecell{$+1.07$ \\$-0.65$}
& $1.02$\footnotesize\makecell{$+0.04$\\ $-0.04$}& $2.03$\footnotesize\makecell{$+0.05$ \\$-0.05$}\\ \hline 
DP+Merge (STFT) &$8.67$\footnotesize\makecell{$+0.98$\\$- 1.27$}&$0.82$\footnotesize\makecell{$+1.92$\\$-1.00$}&$1.75$\footnotesize\makecell{$+0.42$ \\$-0.39$}&$-1.14$\footnotesize\makecell{$+0.36$ \\$-0.89$}&$3.32$\footnotesize\makecell{$+5.95$ \\  $-2.19$}&$-4.57$\footnotesize\makecell{$+4.25$\\ $-4.97$}
& $0.87$\footnotesize\makecell{$+0.04$ \\$-0.04$}& $1.97$\footnotesize\makecell{$+0.65$\\ $-0.30$}\\ \hline 
\end{tabular}
\caption{Comparison of parameter estimation among different data types of DP template in Eq.~\eqref{2p}.}
\label{table2}
\end{table}
\begin{table}[H]
\centering
\begin{tabular}{|c|c|c|c|c|c|c|c|c|} \hline 
Templates &$\Omega_{inf}$ &$\Omega_{PT}$ &$n_1$& $n_{2}$&$n_3$& $n_4$ &$f_1$& $f_2$   \\ \hline  
DP (FFT) & $12.22\%$    & $26.90\%$ & $8.07\%$ & $ 2.31\%     $ &$25.67\%$ & $4.24\% $ & $1.00\%$ & $3.00\%$ \\ \hline 
DP (STFT) &$30.21\%$    & $17.19\%$  & $9.82\%$  & $11.15\%$    & $42.33\%$ & $5.00\%$ & $2.00\%$&$3.00\%$ \\ \hline 
DP+Merge (STFT) &$15.25\%$ & $89.24\%$  & $38.60\%$  & $56.15\%$& $10.67\%$ & $14.25\%$ & $13.00\%$&$3.00\%$ \\ \hline
\end{tabular}
\caption{Comparison of relative errors between the most probable values and the initial values from Eq.~\eqref{parameter} was performed using the DP template in Eq.~\eqref{2p}, while maintaining consistency across different data types.}
\label{table4}
\end{table}

\end{center}
\end{widetext}

\section{Conclusion and Discussion}\label{sec5}

In this paper, we investigate methods for generating mock data from the spectrum and extracting key characteristics from spectrograms in Fig.~\ref{fig4} and Fig.~\ref{fig9}. 
Additionally, we analyze the impact of transient signals on parameter estimation in Fig.~\ref{fig11} and Fig.~\ref{withmerge}, identifying several critical factors that warrant attention during the analysis process. 
These factors contain the influence of noise characteristics, the selection of time-frequency analysis techniques, and the resolvability of transient features within complex signal environments.
It provides a framework for improving signal reconstruction and parameter estimation in the presence of transient and noise-contaminated data, offering a simple example on gravitational wave analysis.

Through analysis of parameter reconstruction methodologies, several conclusions have been obtained. 
When choosing the peak frequency of the DP template in Eq.~\eqref{2p}, the distance between two peaks also has a certain impact on identifying the peak frequency $f_*$. In cases where two peaks are closely spaced in the frequency domain, it becomes challenging to distinguish between them, particularly in the absence of prior knowledge about the spectrum. Under these conditions, the STFT may only indicate that the frequency content is concentrated within a certain frequency range, rather than accurately identifying the precise peak frequencies. This limitation is intrinsic to signal analysis methods that rely on spectrogram, where spectral leakage effects governed by window function choice dominate the visibility of spectral characteristics. To address these challenges, it may be necessary to refine the filtering and analysis techniques. A recent study utilized machine learning methods to separate merged signals \cite{Zhao:2024zkr}, which also provided us with insightful ideas for separating peak signals. Additionally, methods like matched filtering \cite{Owen:1998dk,Kepner:1998pw} and adaptive filtering \cite{Chassande-Mottin:2000zoj,Acernese:2004mp} could further improve the detection and identification of sources in complex backgrounds.

We also propose a processing method of transient signal occurring in the spectrogram and demonstrate its negative influence on the parameter reconstruction of SGWB.
The optimal strategy for mitigating transient signal interference involves their subtraction from synthetic datasets, thereby enhancing analytical robustness and incorporating supplementary prior knowledge. Our result shows that it is particularly important when dealing with strong transient signals, as this effectively minimizes parameter estimation errors arising from such disturbances. Additionally, investigations reveal that transient signals exert negligible influence on spectral reconstruction processes, with their cumulative effects becoming statistically insignificant over extended temporal observation periods.
Transient signals can be obscured by the broader noise spectrum and the superposition of multiple signals, leading to a loss of information during the reconstruction process.
This outcome underscores the challenges of detecting faint transient signals in noisy environments. Even when a signal is visible in the spectrogram, its low amplitude relative to the surrounding noise can lead to significant information loss during spectral reconstruction.
Therefore, if our goal is to extract transient signals, careful selection of time-frequency analysis methods, filtering techniques, and resolution parameters is crucial to ensure that weak transient sources are not obscured by background noise.

The parameter reconstruction analysis in Sec.~\ref{sec4} reveals a critical dependency on template selection. Specifically, the peak-like template exhibits better performance in modeling physical systems characterized by pronounced spectral discontinuities across distinct frequency regimes. Conversely, broken power-law templates demonstrate enhanced applicability to scenarios where the underlying physical processes manifest gradual amplitude modulations over frequency domains. Nevertheless, both templates exhibit inherent limitations in capturing the mechanisms of real gravitational wave generation.
SGWB is not the only scenario that we consider in this research. The real scenario is more complex. Preliminary research on the use of LISA and Taiji to observe some physical models has been carried out extensively~\cite{Auclair:2019wcv,Yao:2024hap,Yu:2024enm,Yao:2024fie,Yu:2023iog,Chen:2024fto,Bartolo:2016ami,Berti:2005qd}. It involves not only cosmological wave sources, such as phase transitions and cosmic strings, but also some new physical models such as Chiral Gravitational Wave Background that may be observable~\cite{Su:2025nkl}. Therefore, SGWB may be intricate. These more complex scenarios will be explored in future research.

\begin{acknowledgments}
This work is supported by the National Key Research and Development Program of China (No. 2023YFC2206200) and the National Natural Science Foundation of China (No.12375059).

\end{acknowledgments}

\appendix

\section{GWs from the Early Universe}
\label{app1}
In this section, we overview the spectrum of GWs from the early universe, including the Inflation and FOPT.

\subsection{GW from Inflation}

Quantum perturbations that occurred during inflation caused tensor metric perturbations, which in turn generated GW \cite{Dufaux:2008dn}.  
Different models and different theories predict different GW sources during the inflation period, such as hybrid, reheating \cite{Turner:1993vb}, kination \cite{Giovannini:1998bp} and particle production \cite{Senatore:2011sp}. Each of these mechanisms has distinct processes for generating GW and predicts different spectral shapes.
We can express $\Omega_{\rm GW}$ in Eq.~(\ref{Omegagw}) as:
\begin{align}
    \Omega_{\rm GW}(f) = C \mathcal{P}_{T}(f),
\end{align}
where $C=\frac{1}{12} \big( \frac{2\pi f}{aH}\big )^2$. 
The spectrum $\mathcal{P}_{T}(f)$ can be written as:
\begin{equation}
    \mathcal{P}_{T}(f) 
    =\mathcal{T}^2(f) \mathcal{P}_{inf}(f),
\end{equation}
where $\mathcal{T}(f)$ is the transfer function \cite{Nakayama:2008wy,Nakayama:2009ce,Kuroyanagi:2011fy,Kuroyanagi:2014nba}. $\mathcal{P}_{inf}(k)$ is the primordial tensor power spectrum, it is commonly parametrized as:
\begin{align}
    \mathcal{P}_{inf}(f) = A_T(f_{*}) \bigg(\frac{f}{f_{*}}\bigg)^{n_{T}}.
\end{align}

For more general case in the inflation, the GW spectrum can be modeled as \cite{Kuroyanagi:2014nba}:
\begin{equation}
\Omega_{\rm GW}(f) =
\begin{aligned}
\begin{cases} 
C\mathcal{T}^2(f) A_T(f_{*}) \big( \frac{f}{f_{*}}\big)^{n_T},  &  f < f_{*}, \\
C\mathcal{T}^2(f) A_T(f_{*}) \big( \frac{f_\alpha}{f_{*}}\big)^{n_T} \big(\frac{f}{f_\alpha}\big)^\alpha, & f > f_{*},
\end{cases}
\end{aligned}
\label{eqA4}
\end{equation}
 where $f_\alpha$ is the frequency at which the power index of the spectrum changes from $n_T$ to $\alpha$. This is the form of peak-like template, we set the $f_{*}$ as peak frequency for power law, where $A_T (f_{*})$ and $n_T$ are the amplitude and the spectral index at $f_{*}$.

\subsection{GW from FOPT}

The GW signal from FOPT has already been calculated intensively. Ref.~\cite{Hogan:1986dsh} and Ref.~\cite{Witten:1984rs} have explored the dependence of the amplitude and characteristic frequency of the GW signal during FOPT.
During FOPT, the parameter $\beta^{-1}$ is used to denote the duration of PT. The precise value of $\beta / H_{*}$ depends on specific models. For instance, if the universe has transitioned to the broken phase, the phase transition must be completed within a time scale shorter than the Hubble time, namely, $\beta/H_*>1$. The typical size of the bubbles $R_{*} \approx v_{b} \beta^{-1}$ , $v_{b}$ is the velocity of the bubble wall.

Due to the coupling between the scalar field and the fluid, sound waves are induced in the surrounding fluid by bubble expansion. Then the sound waves would generate anisotropic stress, which serves as a significant source of GW. Simulations have shown that the sound waves are still a source of GW after the merging of the bubbles is completed. The work of~\cite{Hindmarsh:2013xza} shows that overlapping sound waves, left behind after the transition, are the lasting source of GW. Therefore, the GW sources that last more than a Hubble time are described by a factor $\beta/H_{*}$.

The acoustic production of GW currently lacks a comprehensive model that covers all ranges of parameters $v_{b}$ and $\alpha$, but in the range of values more than about $10\%$ away from the sound speed or the speed of light the results of envelope approximation are consistent with the numerical calculations in~\cite{Hindmarsh:2015qta}. Numerical calculations yield results of the characteristic amplitude:
\begin{equation}
    \Omega_{sw} \simeq 3 \times 10^{-6} \left(\frac{H_{*}}{\beta}\right) \left(\frac{\kappa_{v}\alpha}{1+\alpha}\right)^{2} v_{b},
\end{equation}
where $\kappa_{v}$ is the fraction of latent heat converted into the bulk motion of the fluid.
$\alpha$ is the ratio of the vacuum energy density released in  the radiation bath.

A detailed calculation process for the spectrum have been given in~\cite{Hindmarsh:2017gnf}. The values of $n_1$ and $n_2$ in the SP template \eqref{eq1} can be obtained in ~\cite{Caprini_2016}:
\begin{equation}
    n_1=3, \quad  n_2 = -4. 
\end{equation}
By considering redshift to today, the peak frequency can be expressed as:
\begin{equation}
    f_{sw}\simeq 2\times 10^{-2}{\rm mHz} \left(\frac{1}{v_{b}}\right)\left(\frac{\beta}{H_{*}}\right)\left(\frac{T}{100\text{GeV}}\right).
\end{equation}

\section{The Corner Plots of of Parameter Estimation}

\label{app2}

In this section, we show the corner plots of parameter estimation, based on the SP template in Eq.~(\ref{eq1}), DP template in Eq.~(\ref{2p}),
BPL template in Eq.~\eqref{bpl}.

\begin{figure*}[ht]
\centering
\includegraphics[width=0.5\linewidth]{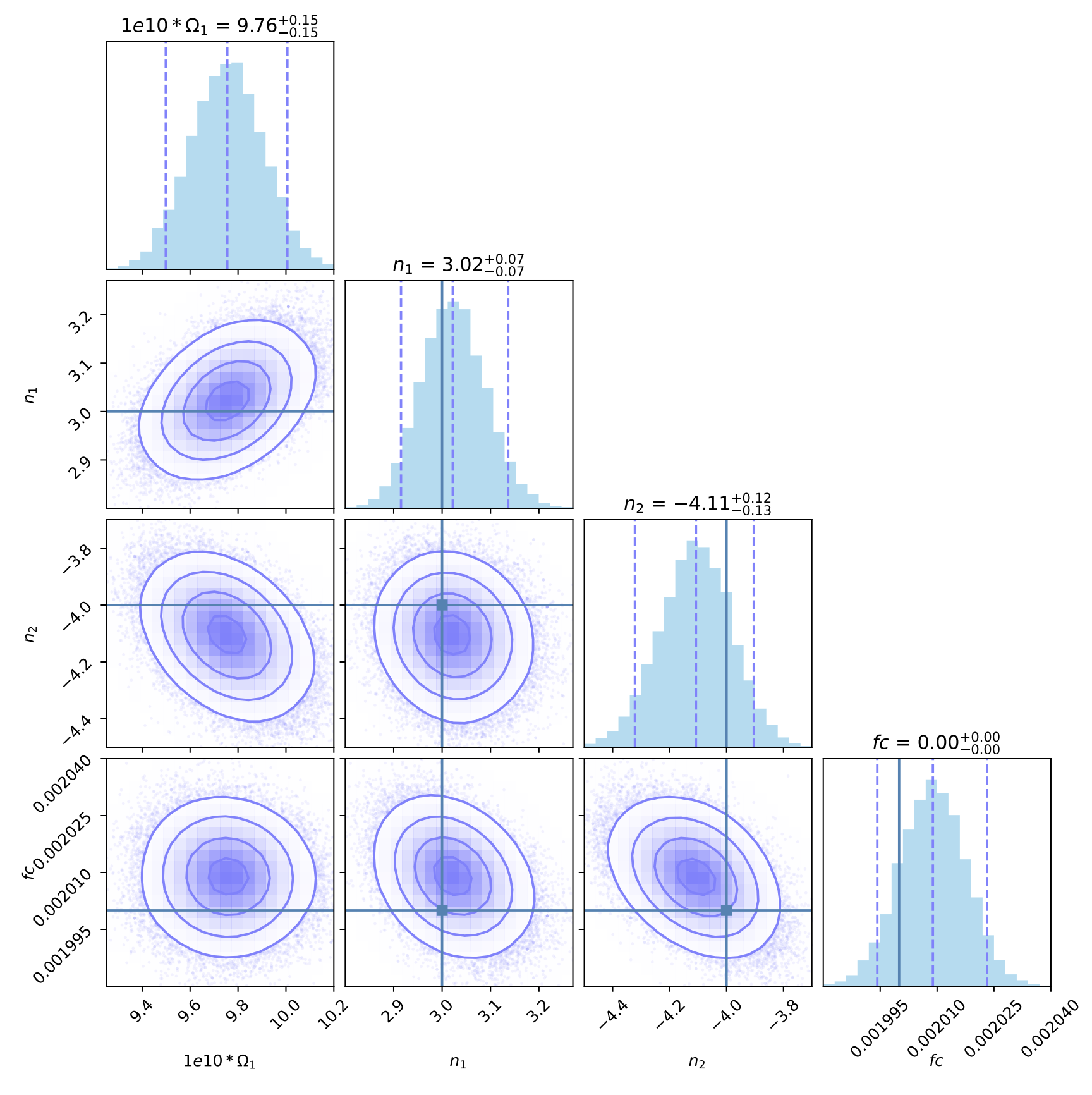}
        \caption{The corner plot of parameter estimation for SP template in Eq.~(\ref{eq1}) by the data from FFT, with parameters in Eq.~\eqref{parametersp}.}
        \label{fig:single}
\end{figure*}

\begin{figure*}[ht]
\centering
\includegraphics[width=0.6\linewidth]{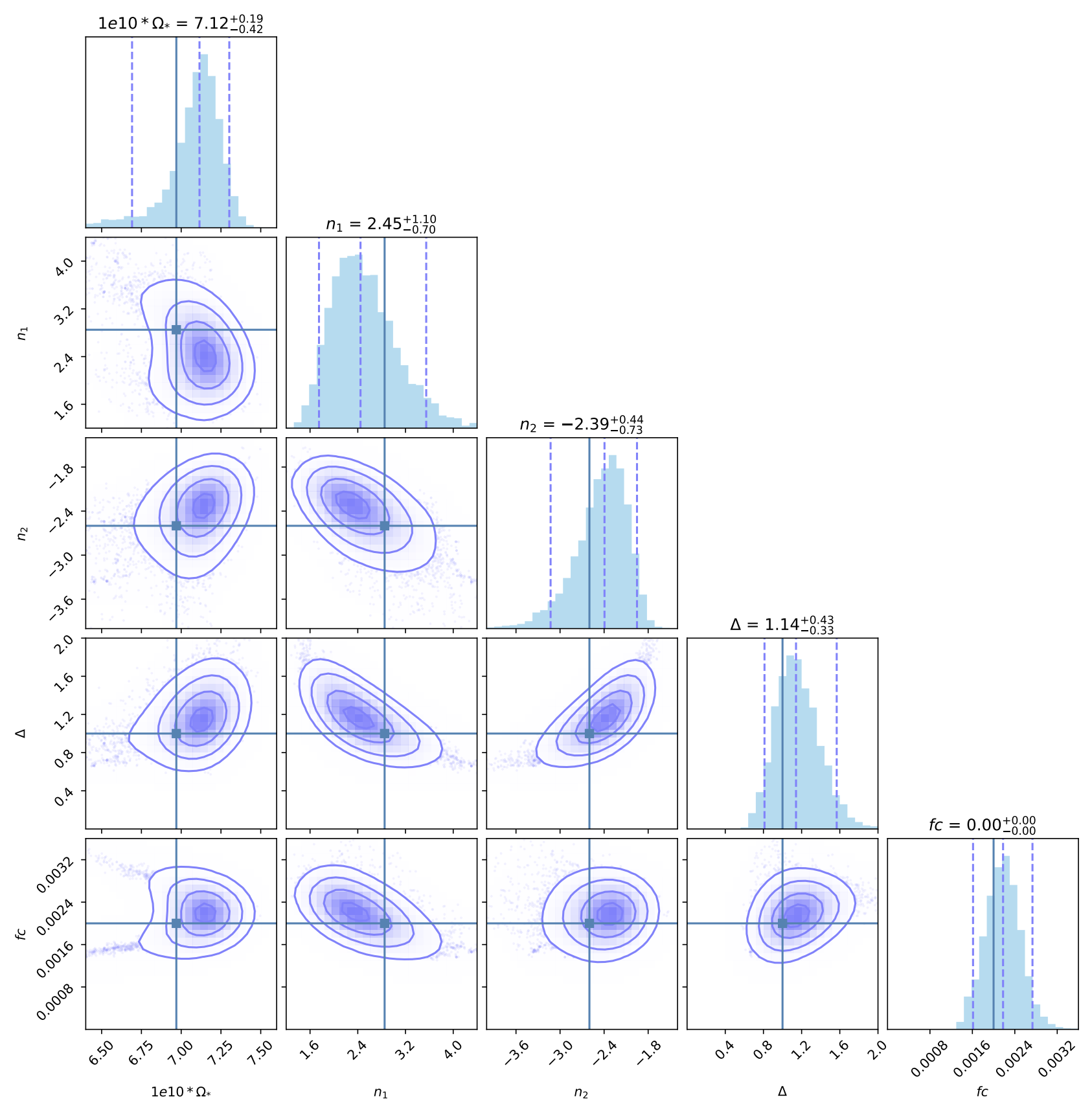}
        \caption{The corner plot of parameter estimation for BPL template in Eq.~\eqref{bpl} by the data from FFT, with parameters in Eq.~\eqref{parameterbpl}.}
        \label{fig:11}
\end{figure*}

\begin{figure*}[ht]
        \centering
\includegraphics[width=0.6\linewidth]{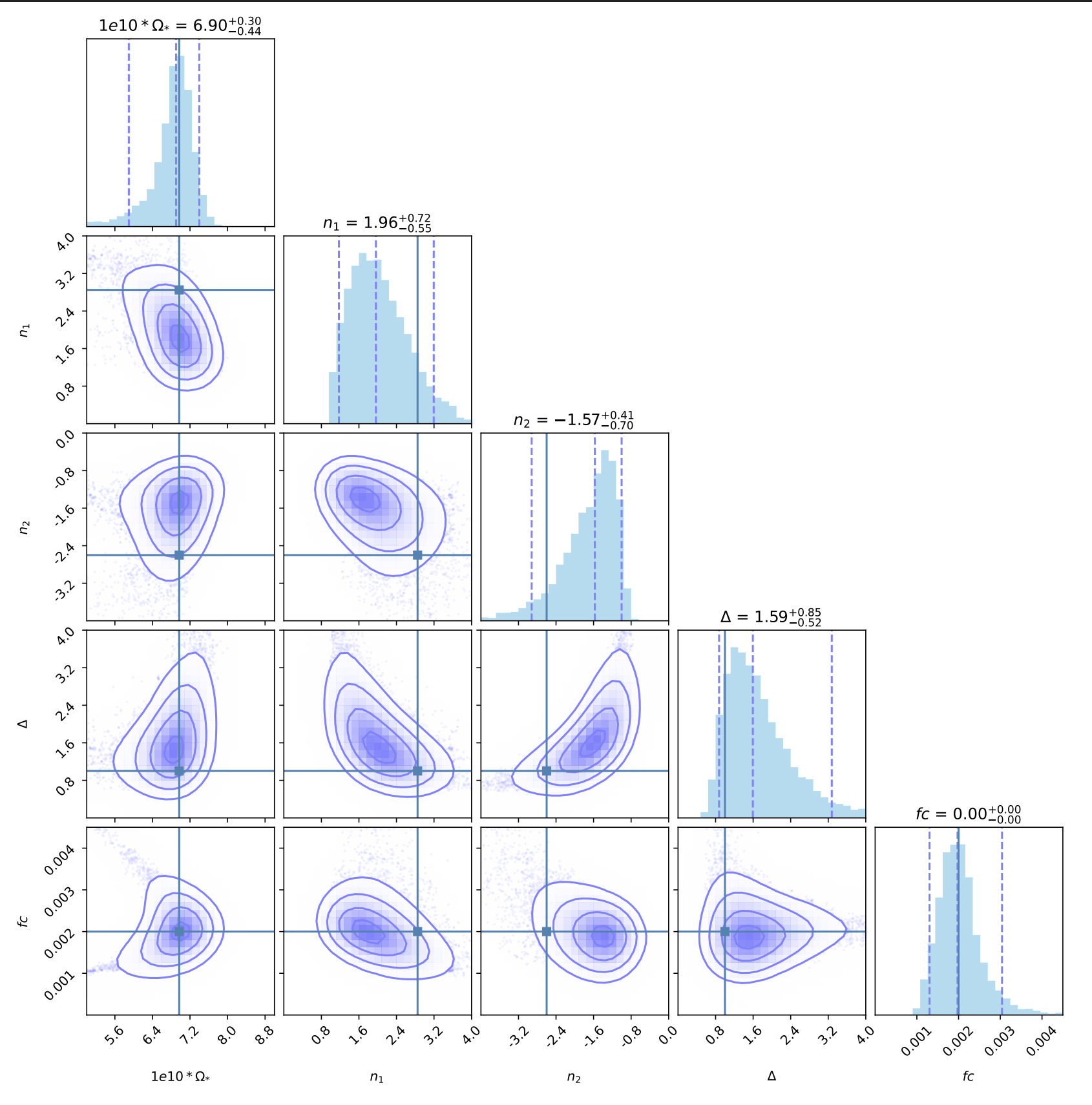}
        \caption{The corner plot of parameter estimation for BPL template in Eq.~\eqref{bpl} by  the data from STFT, with parameters in Eq.~\eqref{parameterbpl}.}
        \label{fig:12}
\end{figure*}

\begin{figure*}[ht]
    \centering
    \includegraphics[width=1\linewidth]{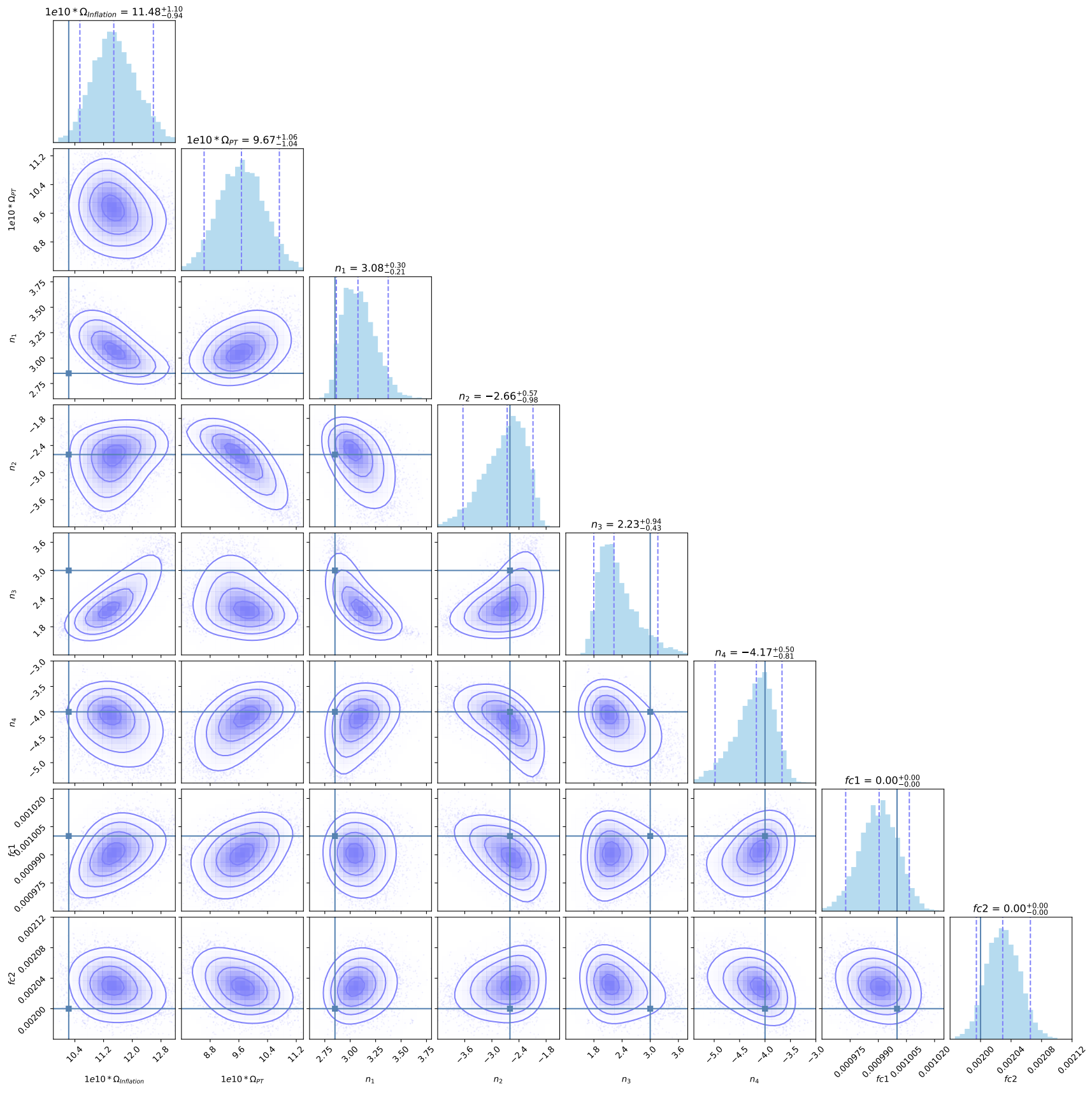}
    \caption{The corner plot of parameter estimation for DP template in Eq.~(\ref{2p}) by the data from FFT, with parameters in Eq.~\eqref{parameter}.}
    \label{fig10}
\end{figure*}

\begin{figure*}[ht]
    \centering
    \includegraphics[width=1\linewidth]{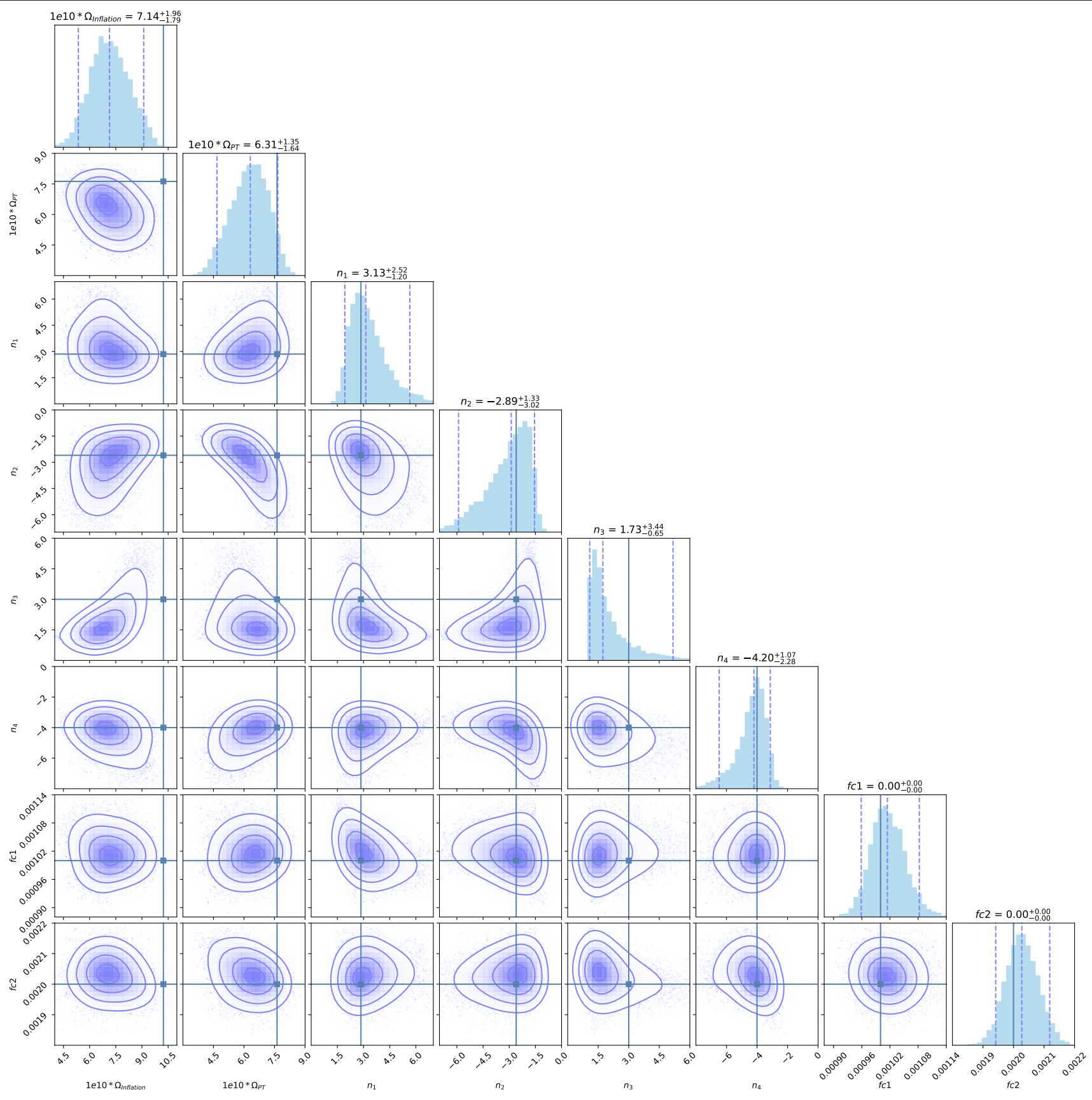}
    \caption{The corner plot of parameter estimation for DP template in Eq.~(\ref{2p}) by the data from STFT, with parameters in Eq.~\eqref{parameter}.}
    \label{fig11}
\end{figure*}

\begin{figure*}[ht]
    \centering
    \includegraphics[width=1\linewidth]{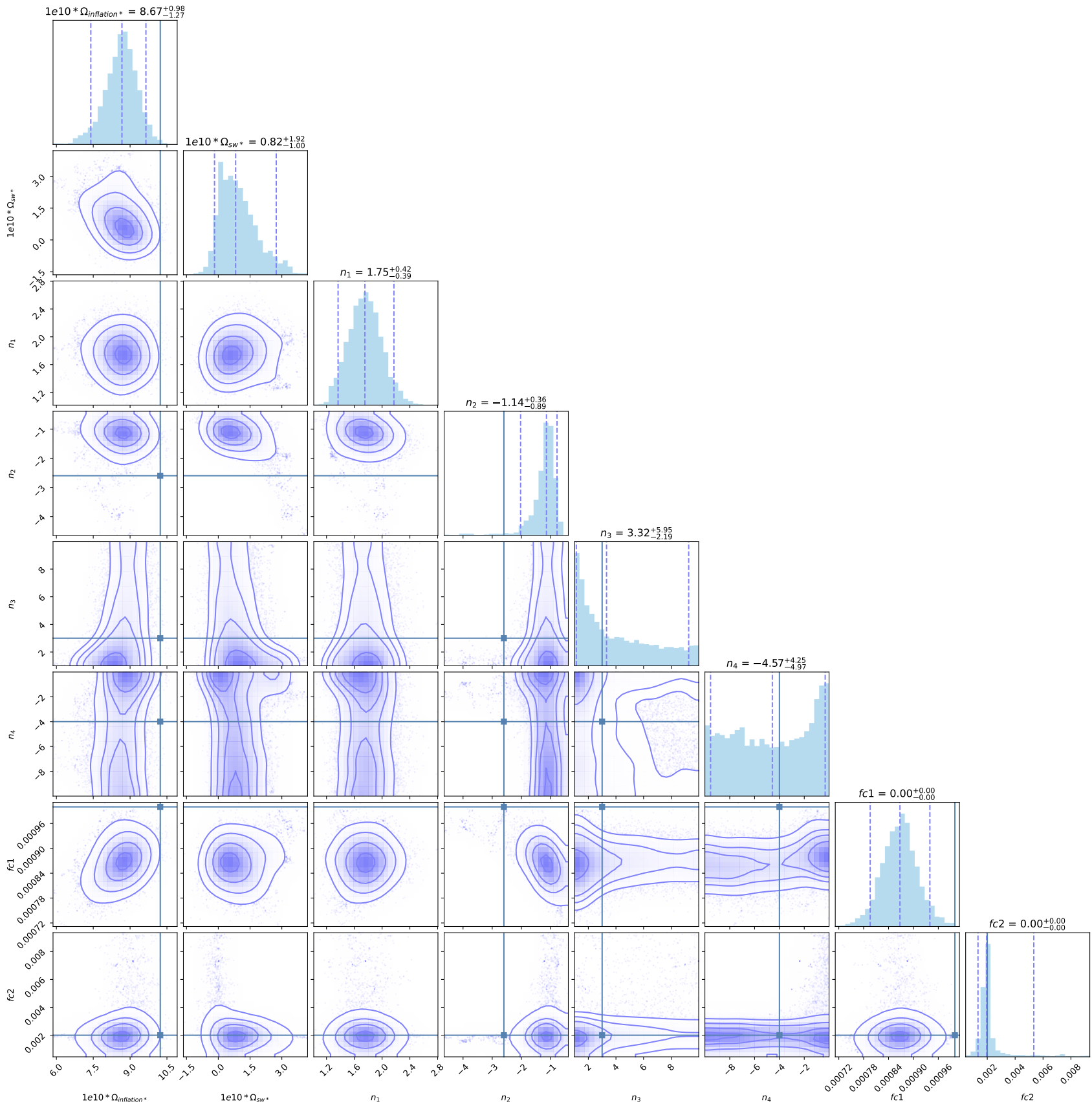}
    \caption{The corner plot of parameter estimation for DP template in Eq.~\eqref{2p} by the data with merge signal from STFT, with parameters in Eq.~\eqref{parameter} and merger signal expression in Eq.~\eqref{merger signal}.}
    \label{withmerge}
\end{figure*}


\bibliography{reference.bib}

\end{document}